\documentclass[graybox]{svmult}

\usepackage{type1cm}        % activate if the above 3 fonts are
\usepackage{makeidx}         % allows index generation
\usepackage{graphicx}        % standard LaTeX graphics tool
\usepackage{multicol}        % used for the two-column index
\usepackage[bottom]{footmisc}% places footnotes at page bottom

\usepackage{newtxtext}       % 
\usepackage{newtxmath}       % selects Times Roman as basic font
\makeindex             % used for the subject index
\begin{document}

\title*{Spacetime is material}
\author{Luciano Combi}
\institute{Luciano Combi \at   Instituto Argentino de Radioastronom\'{\i}a (CONICET; CICPBA), C.C. No. 5, 1894 Villa Elisa, Argentina .\\\email{combi.luciano@gmail.com}}
\maketitle

\begin{abstract}\\
Space and time are central concepts for understanding our World. They are important ingredients at the core of every scientific theory and subject of intense debate in philosophy. Albert Einstein's Special and General theories of Relativity showed that space and time blend in a single entity called spacetime. Even after a century of its conception, many questions about the nature of spacetime remain controversial.  In this chapter, we analyze the ontological status of spacetime from a realistic and materialistic point of view. We start by outlining the well-known controversy between substantivalism and relationalism and the evolution of the debate with the appearance of General Relativity. We analyze how to interpret spacetime as a physical system and how to model its properties in a background-free theory where spacetime itself is dynamical. We discuss the concept of change, energy, and the ontology of spacetime events. In the last section, we review the mereology of spacetime and its relevance in cosmology.

\end{abstract}

\section{Introduction}

Let us assume that the world is made of \textit{things}. Now, assume these things \textit{interact} with each other. From these interactions, some features of these things would \textit{change}. This apparently trivial ontological picture contains three fundamental concepts that have been under dispute since pre-Socratic times: matter, space, and time.

The quest for understanding the nature of space and time has a rich history, both in physics and philosophy. Breakthroughs in science have often shaped and changed the philosophical landscape on this topic, and vice-versa. The standards of the debate were settled in the XVIII century, after the birth of modern theoretical physics by Galileo and Newton. In a famous epistolar debate, Leibniz and Clarke (on behalf of Newton) (Leibniz et al., 2000) discussed the ontological status of space: is space an absolute container where matter exists or is it a concept that arises from the relations among things? (Jammer, 2013). These two opposite views can be broadly classified as \textbf{substantivalism} and \textbf{relationism}. The exact meaning of these two currents of thought and the complexity of the debate have evolved through history, and so precise definitions are strongly context dependent (see Huggett and Hoefer (2018) and Romero (2017)). 

At the beginning of the XX century, Albert Einstein's formulation of Special and General Relativity (GR) produced a profound change in the discussion about space and time. In the words of Weyl:
\begin{quote}
And now, in our time, there has been unloosed a cataclysm which has swept away space, time, and matter hitherto regarded as the firmest pillars of natural science, but only to make place for a view of things of wider scope, and entailing a deeper vision. This revolution was prompted essentially by the thoughts of one man, Albert Einstein. (Weyl, 1922) \bigskip
\end{quote}

General Relativity is a non-linear theory about spacetime and matter\footnote{This implies among other things that spacetime can act as its own source, i.e. it is self-interacting.}. The equations of motion of the theory determine the spacetime metric, which, in turn, represents the notion of distance and inertia for all physical objects. Different from any other previous theory, physical models do not occur in a fixed scenario; GR incorporates the background itself as a dynamical object. Thus, GR shows our three initial concepts harmoniously entangled: space and time constitute a sole entity that interacts with other entities. Famously put by Wheeler:
\begin{quote}
Spacetime tells matter how to move, matter tells spacetime how to curve (Misner et al. 1973).
\end{quote}

Einstein's theory motivated many discussions on the foundations of physics and philosophy in general (Howard, 2014). The radical novelties introduced by GR, and its complexity, pose numerous interpretation problems about its referents. In this work, we will focus on the materialistic aspects of spacetime and relativistic fields. This includes questions such as: is spacetime just another kind of matter field? What is the energy of spacetime? Does spacetime have parts? Does spacetime have entropy? What is a spacetime point?, and other topics. 

First, we offer a sketch of a basic ontological theory that we use throughout the chapter to maintain clarity, and we introduce the main features of relativity in this context. On this basis, we start discussing the substantivalist and relationist positions and their modern formulation. We discuss the nature of events and spacetime properties, emphasizing the background-independent character of the theory. On this basis, we explore different aspects of spacetime as a material being, e.g. spacetime interaction with other systems, evolution, and parts of spacetime. We present the notion of spacetime energy and radiation, and we address its difficulties. Finally, we analyze the main materialistic issues in cosmology.

\section{Matter in Classical Relativity}

\subsection{Basic concepts}

An ontological theory describes the most general features of things. A scientific ontology is an ontology that is systematic, exact, and compatible with updated science (Bunge, 1973). A coherent formulation of such theory can be used as a guide to clarify the most basic concepts with which we deal in our physical theories. A relativistic ontology must encompass in a unified way the description of fields, particles, and spacetime. General Relativity offers a beautiful geometrical formalism for modeling these physical systems. Our job will be to disentangle and understand the basic features of this \textit{relativistic matter} that arise from the theory. On the other hand, making our ontology explicit would help us face fundamental issues about the description of spacetime. We will follow a realistic ontology of things and properties. 

Let us start by stating very basic ontological postulates:
\begin{itemize}

   \item \textit{Things}: A thing is represented as an individual with properties, $X:= \langle x, P(x) \rangle$. Individuals can associate to form new individuals. Properties can be intrinsic (modeled as unary predicates) or relational (n-ary predicates on individuals).
  
   \item \textit{Universe}: The Universe is made of all real things X. The Universe itself is a thing. 
  
   \item \textit{States}: A thing can be modeled as a finite sequence of mathematical functions $\lbrace F_i \rbrace$ over a mathematical space M, where the functions $F_i$ represents a set of properties. A state of a thing is represented as a value of $\mathcal{F}(m)$ for some $m \in M$. 
  
\end{itemize}

We will use these postulates only as a way to drive our discussion. For a more extended development and analysis of this ontological theory see Bunge (1977), Bergliaffa et al. (1993), and Romero (2018). Note that this ontology prioritizes things (primitive stuff) over structure, in contrast with the structural realism of Ladyman (2016), i.e. the idea that no objects are mediating the relations. As we will see in the sections below, GR is suitable to contrast and debate between these two positions (see also Esfeld and Lam (2008)).

\subsection{General invariance and General Relativity}

A physical theory $T$, for a thing $\sigma$ of a certain kind $\Sigma$, has a set of laws that constrains its physical state space (Bunge, 1967). Usually, this is done with a set of differential equations that specify relations between the system and other systems, or with itself. A convenient way to represent the properties, and state-space, of a theory, is using a geometrical language (Thorne and Blandford, 2017). In a geometrical set-up, to formulate physical laws we ought to have a notion of \textit{change} and a notion of \textit{distance} between the geometric objects we define. This structure is called the \textbf{kinematical} framework of the theory (see the superb discussion by Stachel in Ashtekar (2005))

The proper geometric construct to represent this is a metric-affine manifold $(\mathcal{M},\nabla,\mathbf{g})$: the \textit{metric} defines distances and the \textit{affine connection} a covariant derivative, i.e. a way to evaluate changes \footnote{Other geometrical structures are possible for formulating physical theories, e.g. symplectic manifolds for phase space formulation}. This formal affine structure characterizes the \textbf{inertial} behavior of objects in the theory, i.e. the physical state-space of a free system that is not interacting with other systems. A given kinematical framework defines space and time within a given theory (Rovelli, 2018). In Newton's theory, physical distances between objects are Euclidean, sharing a common global time. In Special Relativity, on the contrary, distances are pseudo-Euclidean, meaning that (a) there is no unique global time and (b) there is a causal structure that distinguishes light and massive particles. Both theories share an important feature: the kinematical structure is fixed (Rovelli, 2004). The dynamic of the theory, i.e. the set of equations, occurs on top of this structure. 

Einstein's principle of equivalence (his ``happiest thought'') establishes that inertia and gravitation are of the same ``essence''. This statement was indeed the realization that the gravitational interaction changes the kinematical framework of physical theories and is not an additional force \footnote{This is true even in Newtonian physics. In its geometrical formulation, the affine connection is modified in the presence of gravity with a dynamic Cartan connection (Malament, 2006) while maintaining the Euclidean metric structure.}. In GR, Einstein promotes the entire kinematical structure of Special Relativity to a dynamical structure. This implies a profound \textit{ontological shift} since the kinematical structure, the underlying way to describe the physical laws, is now a representation of a dynamic physical system: spacetime. In short, there are no more fixed structures in the theory, relativity becomes general.

In GR, there is no preferred set of physical (inertial) systems that establishes a preferred reference frame, i.e., there are no privileged inertial coordinates (Rovelli, 2004). The inertial and metric structure is set once we solve the dynamics of spacetime\footnote{Note that these two structures are set by the dynamics of spacetime because the affine connection used is the Levi-Civita connection that is determined by the metric}. The dynamic of GR is given by Einstein's field equations:
\begin{equation}
\mathbf{G}\Big(\mathbf{g}, \nabla \mathbf{g}, \nabla^2 \mathbf{g}\Big) = \frac{8 \pi G}{c^4} \mathbf{T}\Big(\mathbf{g}, \lbrace \phi \rbrace \Big),
\end{equation}
a set of ten second-order non-linear differential equations for the metric $\mathbf{g}$ given a model for the energy-momentum tensor $\mathbf{T}$ of ordinary matter such as fields or particles represented by a set of tensor fields $\phi$. The absence of a fixed kinematical structure implies that the equations are invariant under active diffeomorphism, the most general invertible transformations of a manifold with itself. A model of the theory $M$ is then an equivalence class of these transformations $M= \overline{(\mathcal{M}, \mathbf{g}, \lbrace \phi \rbrace)}$, where $\mathcal{M}$ is the manifold. Diffeomorphism transformations move points in the manifold into other points. If the theory is invariant under Diff($\mathcal{M}$),  we can define the notion of localization only with respect to other physical systems, i.e. properties are always relative to the properties of other systems. This is the core issue of Einstein's Hole Argument that has been well-discussed in the literature over the years (Stachel, 2014). On the other hand, in the classical vacuum case where we have only one entity, spacetime, it can be shown that we still have a consistent and non-trivial theory (see below).

\section{Spacetime ontology}

\subsection{Substantivalism and relationism: the many layers of the debate}
%New addition
Ontological positions about the character of motion, space, and time have been often classified as either \textit{substantivalist} or \textit{relationist}, a distinction most relevant in the physics of the XVIII century that has also been a point of controversy after Einstein presented his General Theory. In order to make sense of the debate nowadays, one has to be careful with the semantic content of the terminology employed; in Rynasiewicz's words:
\begin{quote}
Present day physicists do not employ a language that conforms with the original contrast, and the current controversy is fueled by so much squabbling over how to appropriate modern terminology to one's own doctrinaire advantage (Rynasiewicz, 1996).
\end{quote}

Instead of analyzing what category is more appropriate, let us start from some formal grounds. A precise way to analyze a given theory $\mathcal{T}$ is to take its axiomatic basis $\mathcal{A}$, from where all the consequences (statements) of the theory are derived
\begin{equation}
\mathcal{T} = \lbrace s: \mathcal{A} \rightarrow s \rbrace,
\end{equation}
and specify at this foundational level what we mean by each referential (physical axiom) components (Bunge, 1971). In this way, we characterize what elements are substantial, i.e., \textit{things} in our ontology, or relational, i.e., emergent structures or properties of a collection of things. The concept of spacetime in physics is strongly theory-dependent. Indeed, many discussions have arisen by trying to adapt a given position, e.g., from Newtonian physics, to a more general theory\footnote{By a ``general thing'', we mean a theory $T_1$ whose range of applicability is longer than a theory $T_2$ from where $T_2 \subset T_1$. This implies that the ontology of $T_1$ is more akin to reality.} such as General Relativity (see for instance the dynamical approach by Brown (2005)). A better methodology is to start from the general theory and interpret the restricted theory based on the general ontology,  e.g., from GR to flat spacetime (this is, in fact, non-trivial). In the following, we offer a brief outline of predominant relational and substantial arguments and interpretations in General Relativity. 

One of the most strict forms of relationism about spacetime is a heritage from Mach's ideas into GR. In this \textit{Machian relationism}, matter systems generate the gravitational field that, in turn, determines the inertial properties of every system (see Huggett and Hoefer (2018)). Spacetime then emerges as the system of all the relations between these bodies, i.e., an emergent property of $\mathcal{U}$, the system of all things, and not a thing itself (see Bergliaffa et al. (1998) for a modern exposition of this view). This early interpretation of GR changed after de Sitter presented his solution to Einstein's equation and the famous Einstein-de Sitter debate (Midwinter and Janssen, 2011). This showed that the metric field is not entirely determined by the energy content of other matter, as Einstein expected, although it is constrained by it. It is widely accepted now that the metric represents a physical field, the gravitational field - or spacetime, which interacts with matter and has its dynamics. 

The next question we might pose in the debate is whether spacetime and the gravitational field refer to different things. As we saw in the previous section, there is no independent concept of space and time without the metric. This was noted by Einstein himself and made him change his first Machian ideas:
\begin{quote}
On the basis of the general theory of relativity space as opposed to ``what fills space''' has no separate existence. If we imagine the gravitational field, i.e., the functions $g_{\mu \nu}$ to be removed, there does not remain a space of the type (1) (Minkowski spacetime), but absolutely nothing, and also no ``topological space''. There is no such thing as an empty space, i.e., a space without field. (Einstein, 1956)
\end{quote}

The debate at this point narrows. Note first the substantial and relational character of this interpretation: we have seen that the metric field represents a dynamic entity called spacetime 
(substantial), while the absence of prior background, highlighted by the Hole Argument, renders the theory relational. In the next section, we analyze another important aspect of the debate regarding the substantial character of events.

\subsection{Spacetime events}

Following our sketched ontology, things are represented by elements of a set and properties as functions over these elements. In GR, we can think that each matter field, including spacetime, is an ontological entity. These material systems are represented in the theory by an equivalence class of manifold plus tensor fields. On the other hand, the manifold is itself a set of infinite elements with a topological and differential structure. It is then natural to ask whether manifold points represent some kind of entity with properties (this is what sometimes is called manifold substantivalism, see Hoefer (1996)). Given the background independence of GR, the concept of localization, or spacetime point, is only meaningful after solving the field dynamics. Different from Special Relativity, or Newtonian physics, this implies that matter does not evolve ``on top of anything" and the theory is purely relational between its entities, i.e., the fields. Once the properties of the interacting systems are given, we can distinguish \textit{events}. The ontological status of events is still under discussion, and it has been the main target of the modern substantivalism/relationalism debate (Romero, 2017). An event $e$ is represented as a set of relational statements of point-coincidence properties, e.g., 
\begin{quote}
$e=$ ``spacetime has a Ricci scalar curvature of value $R_0$ when a particle $\sigma$ has a position $X$ and its proper time is $\tau_0$, i.e., $R(X(\tau_0))\equiv R_0$.''
\label{q-e}
\end{quote}

Again, these events are represented by points in an equivalence class of manifolds, once the field dynamic is solved. Each event is associate with a proper \textit{spacetime point}. The question remains of whether we can consistently interpret these spacetime points as things with properties. Suppose that they are, in fact, some kind of factual entities. First, we cannot apply our ontological scheme of Section 1. The main reason is that these spacetime points are not individualized within the theory. We cannot take an event $e$, and formulate the equations of motions for that event because individuation is only inherited from the physical fields: we take the physical fields, we solve the dynamics, and only then it makes sense to talk about the properties, or existence of a particular event \footnote{In this sense, there is a holistic flavor to General Relativity.}. 

Stachel has shown that the picture where the elements of the ontology are the fields is better described in the fiber bundle formalism of GR (Stachel and Iftime, 2005), where we do not need an underlying manifold to formulate the theory. In this approach, he also describes events as having a common nature (what he calls \textit{quiddity}) but lacking individuality (what he calls \textit{haecceity}). In this sense, events seem to have some kind of ontological vagueness that reminds us of similar issues for identical particle states in quantum physics (Stachel et al., 2006).  
 
If events can be ultimately described as things or not will depend on whether a consistent ontological theory of events compatible with GR can be formulated. A clear ontological picture here will lead the way to a theory of quantum gravity  (see Romero (2013, 2016) for further discussions about the equivalence of these two pictures). With these clarifications, we now turn to the problem of change and properties in the context of relativity.

\subsection{Change and properties within spacetime}

A meaningful proposition about the properties of a system in a background independent theory such as GR must be formulated in a gauge-invariant form, i.e. it must be a relational statement. For instance, for a field $\Phi$ with a fundamental scalar property represented by the mathematical field $\phi(x^0,x^1,x^2,x^3)$, we need at least four numbers, or coordinates, to find a complete representation of the property. This complete representation can be built associating these coordinates to the properties of another system that we call reference frame (Rovelli, 2002). We can, in principle, use spacetime itself as a reference frame. If spacetime is non-symmetric, we can establish a set of four unique coordinates $\mathbf{C}=\lbrace \mathcal{C}^{0},\mathcal{C}^{1},\mathcal{C}^{2},\mathcal{C}^{3} \rbrace$, using spacetime properties such as Weyl scalar invariants, which encode the degrees of freedom of the gravitational system. These are known as intrinsic Bergmann-Kommar coordinates (Bergmann and Komar, 1960). A property $\mathcal{P}$ of a given material system can be then represented in general as $P(\mathbf{C})$, where $P$ is a mathematical scalar. This means that \textit{the value of the property $\mathcal{P}$ with respect to $\mathbf{C}$, where spacetime has a unique set of properties, is $P(\mathbf{C})$}. Every local property can be understood to be relational with respect to spacetime: in a well-defined sense, we could say that matter lives on spacetime. In the construction of physical models, however, the use of Bergmann-Kommar coordinates is very far from being a practical choice. Instead, we usually set a coordinate system that has some useful mathematical advantage or external physical interpretation and solve the equations of motion in that gauge. Once this is done, we can build gauge-invariant quantities and extract predictions from the theory. On the other hand, if spacetime has symmetries, then we need material systems acting as reference frame since we cannot distinguish certain states or parts from others using only spacetime properties (Smolin, 2006).

Although spacetime is always interacting with other fields, spacetime can be dynamic even in the absence of other material entities, e.g. black holes can merge in a vacuum. This poses an obvious concern: how spacetime can be dynamic---changing in some sense---if there is no other thing to compare these changes with? One could be tempted to say that in vacuum there is no dynamics (or  that ``time is frozen''), as it is mentioned by some authors. But this would be against the usual practices of physics, and in fact, incorrect. The reason for this confusion is that there is no external time in GR, in contrast with Newtonian theories. Moreover, as we mentioned, other conceptual difficulties arise when we try to formulate well-behaved invariant quantities in a diffeomorphism invariant theory. In this section, we will focus on the philosophical part of the discussion, while the reader is invited to read Pons and Salisbury (2005) for the technical issues.

Spacetime properties are four-dimensional. We can fully characterize these properties, e.g. by putting them in terms of Bergmann-Kommar coordinates. We can then separate or 'slice' these properties choosing an arbitrary foliation of space-like surfaces $\Sigma$, and a perpendicular time direction\footnote{This is possible assuming that the manifold is well behaved globally}. In this 3+1 formulation, the evolution of spacetime can be described as the evolution of one 'chunk' with respect to an intrinsic time. If there is a global choice of time-variable $t$, we can parametrize each state as $\Sigma(t)$. This is the usual way to analyze and solve numerically models of high complexity in General Relativity (Baumgarte and Shapiro, 2010). On the other hand, spacetime properties are partially ordered. i.e. there is an equivalence class of time-like curves connecting these properties. In the 3+1 picture, we are choosing a collection of spacetime properties ordered in some fashion. An asymmetric relation between these ordered properties shows that spacetime is non-trivial and dynamic. This asymmetry between the past and future is an invariant fact of spacetime, but the concrete description of the evolution depends on the frame we choose \footnote{Paraphrasing Pierre Curie, \textit{l'asym\'etrie cr\'ee le ph\'enom\`ene}}.

It is important to note that spacetime is not the physical system of all the foliations $\lbrace \Sigma(t) \rbrace$. A physical system is the composition of two physical things, e.g., a hydrogen atom whiich is made of an electron and a proton. In this case, each foliation is a given state that we choose to describe the evolution of spacetime properties. Each of these foliations does not have an independent ontological status by itself. Although we can distinguish between different spacetime states, what is impossible to determine in the absence of other physical systems is an intrinsic length scale between these two events, i.e., it is not possible to measure a rate of change. Moreover, to establish an actual scale on spacetime, a concrete parametrized time-like curve is needed. This implies that a scale (e.g., clocks and rods) can only be established by massive systems, i.e., a system with non-zero energy in its fundamental state. 

If an entity is interacting with spacetime, we can represent changes with respect to a reference frame attached to this thing (see Khavkine (2015) for self-consistent toy models that achieve this). Nevertheless, a formulation of change within spacetime is important to the extent that spacetime is the sole entity that can haven non-trivial properties, at least classically, in the absence of other material systems. Any other system is subject to the interaction with spacetime. This is implicitly stated in the formulation of physical theories. Summing up, we can state that spacetime is a changing entity but not with respect to an external time, as in a Newtonian ontology. Its change is always measured against other material systems or, in a weaker sense, we can define change noticing that the properties of spacetime are ordered.

Finally, it is useful to distinguish two important notions that are often seen as complementary: \textit{existence} and \textit{becoming}. Existence is a concept that cannot depend on conventions. If something exists, we must describe it as an invariant object of the theory. In General Relativity, this is best seen through a global conception of existence: a material system exist as a 4-dimensional ``worm" \textit{occupying} a hypersurface $\mathcal{D}$ (Heller, 1990), i.e., the spacetime region where the system interacts. In other words, the existence of a system $\sigma$ is associated with its entire state-space, which defines a spacetime domain $\mathcal{D}$ inheriting, from spacetime, a casual structure on it (Malament, 2006). Usually, when we ask what \textit{is} an object, we refer to this structure. In a Parmenidean sense, since this state cannot change, relativity commits to Eternalism (Romero, 2013). 

Our ontology postulates agree with this view since a concrete thing is defined as an individual with the entire state space of its properties. Although the existence of matter is unchangeable as a 4D object, this does not mean that time or change is altogether an illusion for entities. The important point is that General Relativity describes changes between systems as an objective fact of nature, but there is no preferred time variable to describe these changes. Contrary to a Newtonian ontology, what \textit{exist} is not attached to a particular moment. On the other hand, note that physical laws are concerned with local change. The language of events or becoming seems more natural for doing physics; to test our theories, we look for processes and events. But events mark the interaction between two things. As we said, a pure ontology of events would have to look essentially different from our thing-based ontology.

\subsection{Matter on spacetime}

In this section, we take a closer look into fields and particles, and their relation to spacetime (see also chapter by Romero in this volume). A field $\Phi$ is an entity with a fundamental relational property (or several of them) that we call intensity, represented as a tensor field $\phi(\mathbf{X})$, where $\mathbf{X}$ are four-dimensional coordinates. On the other hand, a particle $\gamma$ is an entity with a fundamental relational property called momentum, represented as a one-dimensional vector field $\mathbf{p}(\mathbf{X})$. Field and particles can also have intrinsic properties such as mass and charge that are imprinted in the equations of motion. Both fields and particles have an important property called energy-momentum. Energy, as an ontological property, represents the changeability of a thing (Bunge 2010). Every physical theory has a representation of energy for its physical objects. Noether's theorem beautifully connects the energy of a system with the properties of spacetime. Given a time symmetry in spacetime, represented by a Killing vector $\mathbf{t}$, if the state of a system is invariant with respect to $\mathbf{t}$, then the energy of the system is conserved. In general, we do not have a preferred notion of time in spacetime, so conserved quantities only arise in very special cases.

Energy-momentum is the relativistic extension of the concept of energy. In GR, this is represented as a symmetric tensor $T_{\mu \nu}$ which is defined as the functional derivative of the Lagrangian $\mathcal{L}$ with respect to the metric:
\begin{equation}
T_{\mu \nu} := -2 \frac{\partial \mathcal{L}}{\partial g^{\mu \nu}} +\mathcal{L} g_{\mu \nu}.
\end{equation}

From this definition, we observe that the concept of energy-momentum is linked intrinsically to spacetime since the energy-momentum tensor depends explicitly on the metric $\mathbf{g}$. As stated by Lehmkuhl (2011), the concept of energy cannot be conceived without an underlying of spacetime. On the other hand, the dynamics of spacetime given by Einstein's equations are only affected by matter through its energy-momentum. This can be understood as a consequence of the strong equivalence principle (De Felice and Clarke 1992): no matter the intrinsic nature of matter, spacetime is only concerned with its energy-momentum.

Particles are part of most classical ontologies both in Newtonian and Relativistic physics. Their interaction with spacetime is through the inertial structure, determined in GR by the metric. The dynamic of a particle is given by the force equation:
\begin{equation}
\mathbf{p} \cdot \nabla \mathbf{p} = \mathbf{F},
\end{equation}
where the covariant derivative $\nabla$ is determined by the affine connection and $\mathbf{F}$ is the force, denoting the interaction with other entities.Contrary to a field, a particle cannot interact with itself and its energy-momentum encodes a complete representation of its properties \footnote{The energy-momentum of a particle is not well-defined, but we can build a tensor in the sense of distributions.}. This is why the state of the particle is completely determined by the conservation of the energy-momentum. In the absence of forces, a particle only interacts with spacetime. If we consider the back-reaction of the particle onto spacetime in this idealized model, we get an ill-defined metric that far from the particle position, looks like the Schwarzschild metric of mass $m$ (Katanaev, 2013). Physically, however, we know that any sufficiently dense distribution of mass would lead, in the end, to a black hole, and so this model is only valid at certain scales. Indeed, there are several reasons to think that particles are not fundamental in the ontology of the World. In the quantum regime, particles are frame-dependent excitations of some underlying field. In the classical regime, we can also think of particles as the geometrical optic limit of some field, where the wave behavior of the latter is washed-out at some limit (Misner et al. 1973). The intrinsic properties of the field would determine the initial inertia state of the particle; if the field is massive, then the particle moves in a time-like curve, while a massless field follows null curves.

If we consider an average description of a collection of non-interacting particles we can formulate the simplest field theory: a dust fluid. This field is characterized by a rest-mass density field $\rho(\mathbf{X})$ and a velocity field. Similarly to a single particle, the dynamic of a dust field is given by the conservation of the energy-momentum alone. If we include interactions into the field model, we need to add other fundamental quantities such as pressure and viscosity (Thorne and Blandford, 2018). Contrary to a fluid, which is a statistical description of more fundamental entities, there are fields in nature that are fundamental. In the classic regime, the most relevant example is the electromagnetic field, with a fundamental property represented by the Faraday tensor $F^{\mu \nu}$. Since the electromagnetic field is an unobservable entity, there have been debates about its physical reality (Bunge 1967) and attempts to formulate theories without them. It is now widely accepted that fields are real referents of our physical theories for various reasons, including most notably their successful description of physical reality in the quantum realm.

\subsection{Gravitational energy, radiation and thermodynamics}

As we discussed above, the energy of matter is closely related to the notion of change in spacetime. The absence of an external (or preferred) time variable in spacetime makes the problem of assigning energy to gravity non-trivial. By the equivalence principle, the local properties of a free system that moves in curved spacetime remain unchanged. In other words, locally, spacetime looks flat. This shows that a local notion of gravitational energy is elusive. Gravitational energy must be, at least, a quasi-local concept, a quantity that characterizes an extended region and not a local region. Besides, to build physically relevant energy associated with spacetime, we have to be careful and described it for a specific reference frame.

A natural choice is to separate two regions of spacetime, one that is highly dynamical, $\mathcal{D}$, and another far away, $\mathcal{O}$, where nothing happens. In this manner, we try to quantify the energy in $\mathcal{D}$ with respect to $\mathcal{O}$. This is the basic idea of the ADM (Arnowits-Desser-Misner) mass, $M$, defined as the surface integral:
\begin{equation}
M = -\frac{1}{8 \pi} \lim_{r \rightarrow \infty} \oint_S (k -k_0) dS,
\end{equation}
where $k$ is the extrinsic curvature of a surface $S$, $k_0$ is the extrinsic curvature of $S$ embedded in flat spacetime, and $dS$ is the surface element (Poisson 2004).  This quantifies, from the superficial curvature, the energy contained in the inner region, as described by a far away observer. When we speak about the mass of black holes, we are usually referring to this construct.

In a dynamical scenario like the inspiralling and merging of two black holes, part of the energy is radiated away from the dynamical region, and the rest remains in the remaining black hole. The final mass of the black hole cannot be calculated using the ADM construct, because the prescription to calculate the energy includes all the gravitational energy in the domain, the radiation and the stationary mass. In this case, to calculate the remaining mass, we have to use a concept of energy that takes into account this radiation separately. The useful quantity here is the Bondi-Sach mass, $M_{BS}$, which uses a null-surface to measure the energy, meaning that any radiated energy remains in the past. The rate of change with respect to a locally flat time is given by:
\begin{equation}
\frac{d M_{BS}}{du} = - \oint_{\mathcal{I}^+} F dS,
\end{equation}
where $F$ is the gravitational flux (see Figure 1 to compare both ADM and Bondi-Sachs constructs). Contrary to the ADM mass, this mass can be connected directly with observations at a gravitational wave detector by measuring the energy flux, and thus infer the final mass of the black hole.

\begin{figure}[t]
\sidecaption[t]
\includegraphics[scale=.45]{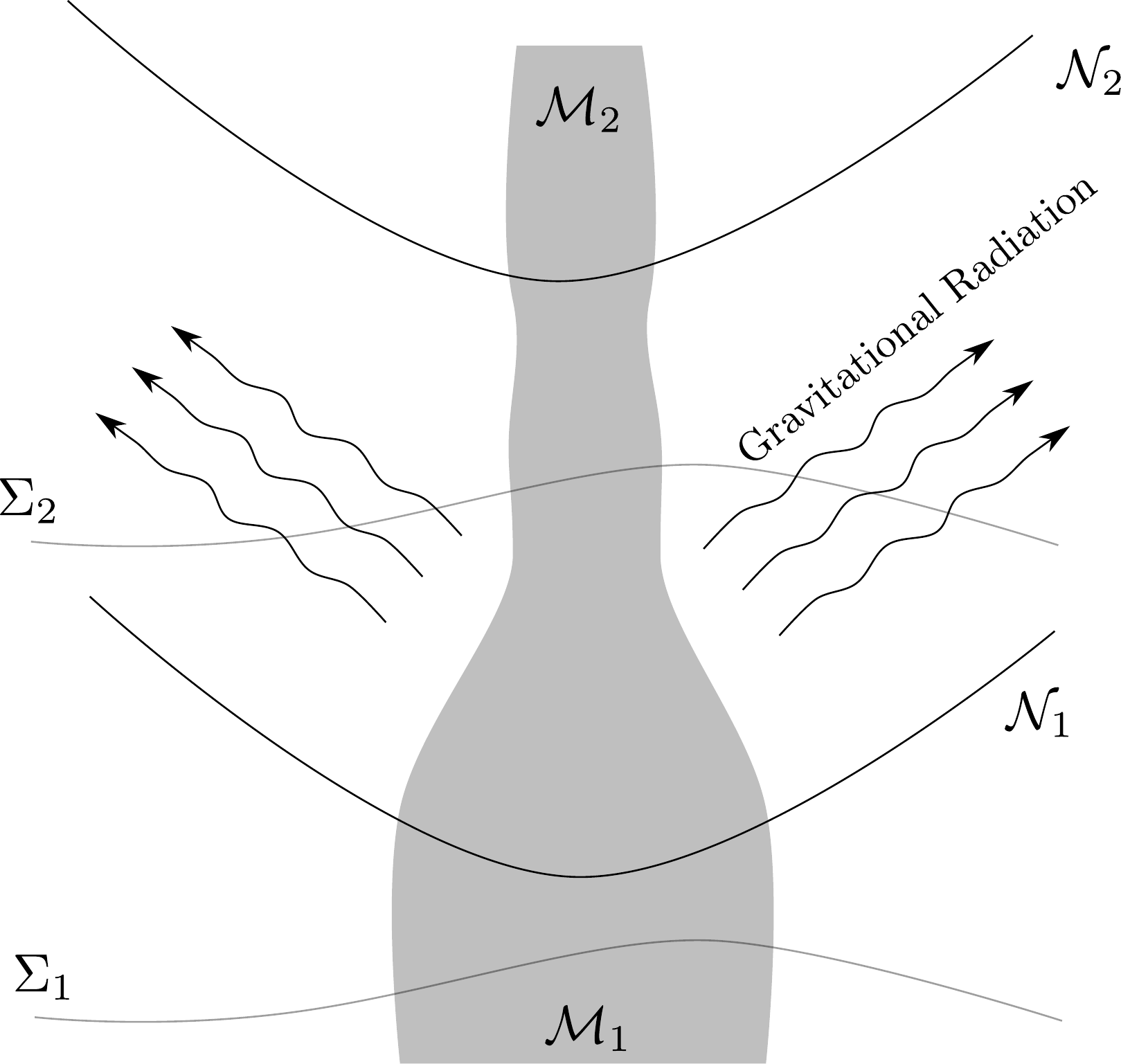}
%
% If no graphics program available, insert a blank space i.e. use
%\picplace{5cm}{2cm} % Give the correct figure height and width in cm
\caption{In this figure we show a sketch of a gravitational source, for instances, a perturbed black hole, with mass $\mathcal{M}_1$ at an initial time as computed with both the ADM and Bondi-Sachs mass. After the perturbations are radiated away as gravitational waves, the rest mass of the black hole is $\mathcal{M}_2$. The ADM mass contained in the spatial surface $\Sigma_2$, however gives us information about all the energy available in space-time so again it gives $\mathcal{M}_1$. On the other hand the Bondi-Sachs mass, computed with a null surface $\mathcal{N}_2$ leaves the radiated energy in the pass, giving the proper mass of the black hole.}
\label{fig:1}       % Give a unique label
\end{figure}

The previous energy concepts we showed need a reference frame at infinity, assuming that very far from the source, spacetime is flat. This precise notion allows to properly characterize an  \textit{isolated system} in curved spacetime (Wald 1984). Instead, if a given system is embedded in an expanding Universe, these definitions need to be revised (see next section). The energy constructs that we described so far are global in character. In principle, we could apply the same reasoning in a \textit{quasi-local} manner, that is, in finite region around a certain point. For the quasi-local case, we do not have a preferred way to choose our reference frame so there are many quasi-local energy definitions. Two of the most mathematically well-defined and physically motivated concepts of energy are the Hawking and Brown-York energy formalisms. Another alternative is to try building an energy-momentum for spacetime. These are usually pseudo-tensors (Landau, 2013) and, again, are only physically meaningful when a proper reference frame or a set of physical coordinates is chosen.

The fundamental character of spacetime in defining the kinematical structure of matter has interesting consequences in thermodynamics for two reasons: the formulation of the first law and equilibrium in curved spacetime, and the role of spacetime in increasing the entropy of a system. For instance, tt can be shown that the interaction of spacetime and matter in a thermal bath can create temperature gradients in equilibrium without violating the first law of thermodynamics, e.g. in presence of a black hole we can have the Tolman temperature for a stationary fluid (see Santiago and Visser (2018) for a recent discussion on this topic). 

The question of entropy and the second law is more subtle. If the second law of thermodynamics is globally valid, then there seems to be a contradiction on how the Universe evolved from a thermal equilibrium state that maximizes the entropy and the current state of the Universe that is colder and structured. A usual proposed solution to this problem is to assume that spacetime carries entropy itself and in the early Universe it was in a low entropy state. Then, via gravitational collapse, spacetime entropy grows with the formation of black holes as a limiting case (see Figure 2).  Although we do not have a microscopic theory of spacetime to obtain a self-consistent measure of entropy, there have been attempts to find prescriptions for a classical entropy. A promising proposal is due to Penrose, who proposed that the conformal invariant Weyl tensor, representing the degrees of freedom of spacetime, could be used to build a proper estimation of the entropy (see Clifton et al. (2013) for a concrete realization). This is also motivated by the second law analog of entropy found for black holes, where the entropy of a black hole is proportional to its area, which is always a growing (Bekenstein 1973). 

This simplified picture of the thermodynamics of baryonic matter and spacetime has some nuances (Wallace 2010). First, note that spacetime acts mainly as a catalyst for entropy increasing processes, clustering matter to allowing thermonuclear reactions in stars and other systems. The process of clustering in the presence of gravity indeed can be entropically favored since, although matter entropy decreases when a structure is formed, it emits highly-entropic radiation. In most local scenarios, e.g. to develop life on earth, the gravitational entropy itself is not relevant to increase the local entropy; in the cosmological picture, however, black holes carry an enormous amount of entropy. In this context, it is argued by Wallace that the low entropy state of spacetime in the early Universe is not the only reason why the Universe is evolving to a more entropic state; in short: (a) when gravity is taken into account, non-uniformity increases entropy via thermal processes induced by gravity, and in particular cases by black holes, and (b) the Universe was not in global equilibrium due to the rapidness of its expansion. Even though these remarks are important, it is clear that spacetime stores entropy and energy in a very different way than normal matter.

\begin{figure}[t]
\center
%\sidecaption[t]
\includegraphics[scale=.45]{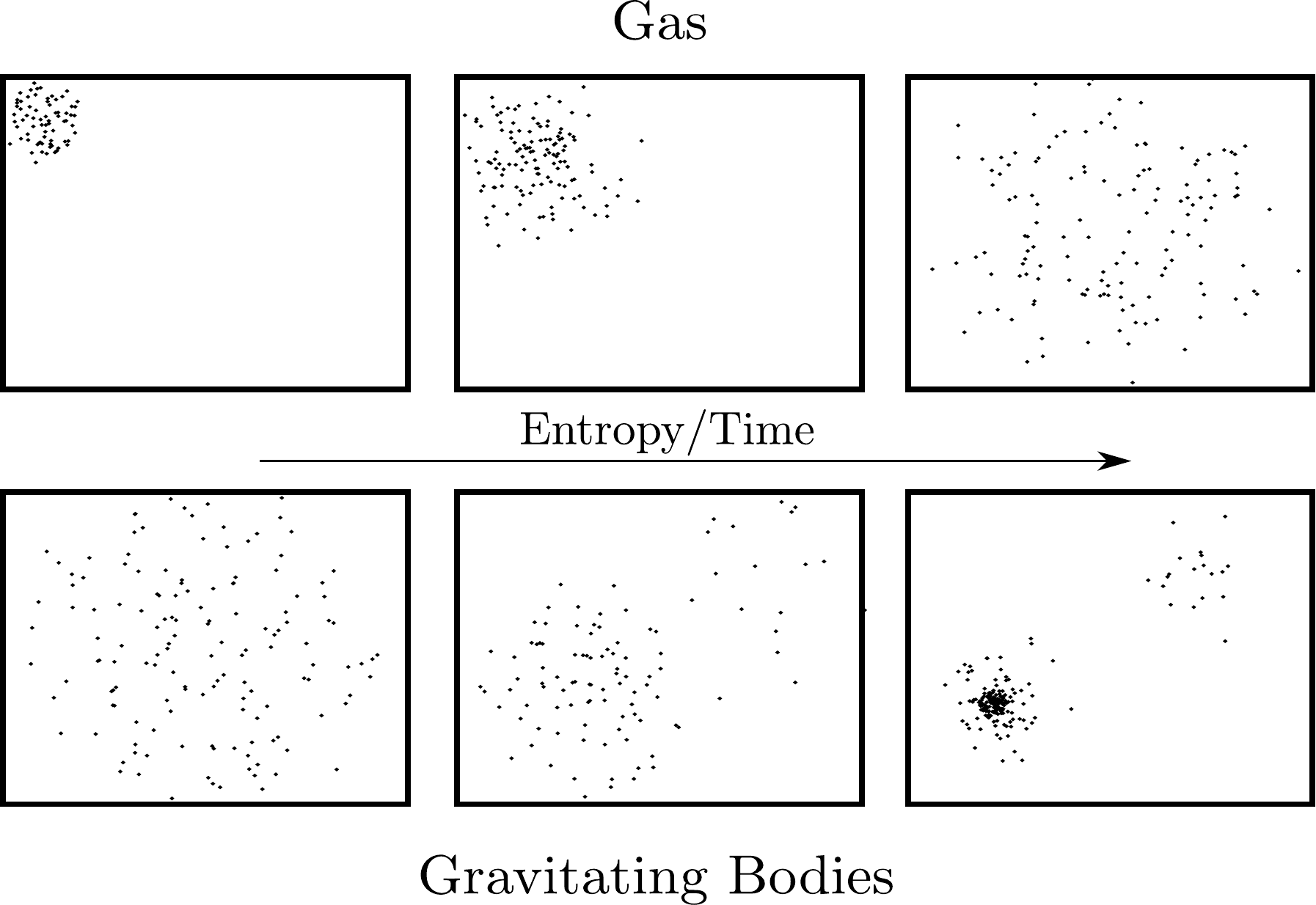}
%
% If no graphics program available, insert a blank space i.e. use
%\picplace{5cm}{2cm} % Give the correct figure height and width in cm
\caption{Thermodynamic picture of a system without (up) and with (down) gravitational interaction taken into account.}
\label{fig:1}       % Give a unique label
\end{figure}

\section{Cosmology and the scales of spacetime}

Cosmology, the study of the Universe as a whole, started as a scientific discipline with Einstein's first model of a static Universe in 1917 and the discovery of the redshift-distance relation of Hubble, showing that the Universe is dynamic and expanding. Although cosmology has always been (and to some degree is) a concern for philosophers, nowadays, with large surveys and advanced telescopes, cosmology has become a precision science. The standard model of cosmology is the so-called $\Lambda$CDM model (Ellis et al. 2012), that most notably implies that (a) the Universe is homegenous and isotropic at large scales, (b) is currently expanding in an accelerated manner, (c) there was a Hot Big Bang in the past, and that (d) equations of GR must be supplemented by a linear term with a small constant $\Lambda$ called the Cosmological Constant.

The philosophical issues around cosmology are vast and it could take an entire chapter (see the excellent discussion of Ellis in Butterfield and Earman (2006)). But here we are concerned with the material aspects of the Universe. In particular, with the following three issues: is the Cosmological Constant a thing? what are the parts of an expanding Universe? how do these parts interact?

Let us begin by the nature of the Cosmological Constant. First introduced by Einstein, the cosmological constant is the simplest modification of Einstein's equations in the form:
\begin{equation}
\mathbf{G}(\mathbf{g}, \nabla \mathbf{g}, \nabla^2 \mathbf{g}) + \Lambda \mathbf{g} =  \frac{8 \pi G}{c^4} \mathbf{T}(\mathbf{g}, \lbrace \phi \rbrace ),
\end{equation}
Often referred as ``dark energy'', the $\Lambda$ contribution plays a fundamental role in modern cosmology since it provides an explanation for the accelerated state of the Universe (Carroll 2001). Physicists often speak about a fundamental mystery surrounding this constant. Although this has motivated alternative explanation for $\Lambda$ and new theories, the character of the problem is greatly exaggerated (see Bianchi and Rovelli (2010)). There are two possible ways to interpret the constant, either (a) we reify it or (b) we accept that we must modify Einstein's equations. Its reification, promoting Lambda to a physical property of some entity, comes in two flavors. First, we could assume that $\Lambda$ represents a property of spacetime itself, the vacuum energy. It is a well-known problem that the calculation of the vacuum energy performed in Quantum Field Theory gives a result of 55 orders of magnitude larger than the actual observed value. As it is explained in Bianchi and Rovelli (2010), it is a mistake to a prior identify this constant as the vacuum energy, since the mechanism of quantum gravity is still unknown. Second, we could assume that $\Lambda$ is a property of a quintessential substance that is dynamic, e.g. a scalar field. No compelling evidence has been found to prove this so far. Although there are interesting alternatives for the Cosmological Constant such as $f(R)$ theories, there are no convincing arguments about why $\Lambda$ should not be just a fundamental constant of nature such as $G$ and $c$ that is included in the equations. The fuzz is indeed unjustified.

In any case, the presence of the Cosmological Constant changes the structure of spacetime itself.  When spacetime does not possess asymptotic flatness, it is hard to extract information of isolated, e.g. energy measures and gravitational waves (Ashtekar et al. 2014). Asmyptotic flatness endows spacetime with several desired properties that are not present when the cosmological constant is present \footnote{In particular, when $\Lambda \neq 0$, the well-defined notion of null boundary at null infinity is a spacelike surface and thus there is no preferred way to calculate physical quantities. When spacetime is expanding all notions of infinity are origin-dependent; natural boundaries appear in spacetime known as horizons (Ashtekar and Krishnan 2004).}.

Because of the non-linear nature of spacetime,  the interaction between small scales and large scales is non-trivial (see Figure 3).  There are indeed interesting \textbf{top-down} and \textbf{bottom-up} problems. Let us revise the bottom-up problem first. In the early universe, the matter was in a hot and dense plasma state. This plasma is very well-approximated \textit{everywhere} by the energy-momentum tensor of a homogenous and isotropic perfect fluid. The solution to Einstein's equations for this case is the well-known FLRW metric. As the universe evolves and structure is formed, the matter is homogeneous only over big scales of hundreds of megaparsecs. To solve the dynamic of the whole universe we thus have to take averages on these scales and solve the averaged  Einstein's equations. If the averaged matter is homogenous, do we recover the FLRW metric? It turns out that given the non-linearity of Einstein's equations, the averaged equations are different from the exact equations. This is the so-called backreaction problem: does the presence of inhomogeneities change the global solution?

An important debate among cosmologists has been going on along the past decade to address this problem. Green and Wald (2011) have argued that in a weak-perturbation scheme where the dynamics of the Universe is described a metric with the parts:
\begin{equation}
g_{\mu \nu} = g_{\mu \nu}(0) + \gamma_{\mu \nu},
\end{equation}
being $g(0)$ the homogenous Universe and $\gamma$ representing the arguably small inhomogeneities, backreactions are very small and not relevant in observations. This is not an obvious conclusion since derivatives of the perturbations are not bound by how small the perturbations are. This is the important result demonstrated by Green and Wald (2011). A strong response to this work has been given by Buchert et al. (2015), where they express the limitation of the Green and Wald scheme (see also the response by Green and Wald (2015)). In particular, they point out that the way they consider backreactions is not general. In the Green and Wald scheme, local deviations to the homogenous spacetime are small with respect to the whole global background. If instead, we consider small ripples that accumulate over the averaging at large scales, it is possible to build a global inhomogeneous model of the Universe (Korzynski, 2015). Inhomogenous cosmologies are relevant because they could explain current observations of the accelerated Universe as an alternative to the standard model (Buchert et al. 2016).

\begin{figure}[t]
\sidecaption[t]
%\sidecaption[t]
\includegraphics[width=0.6\textwidth]{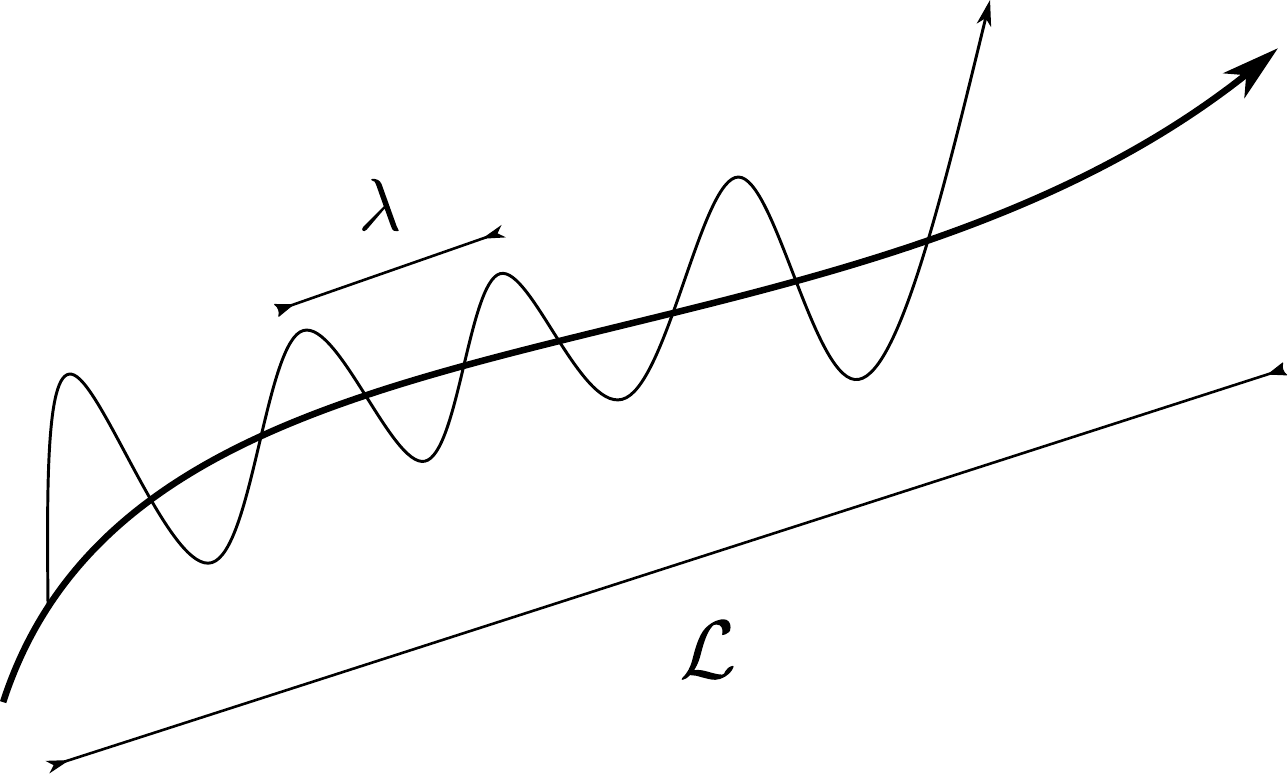}
\caption{In some cases, spacetime properties can be separated into a background, with some scale $\mathcal{L}$, plus perturbations, with a scale given by $\lambda$. These are also known as the low-frequency and high-frequency modes, respectively. This is a tipical scenario for gravitational waves or cosmological perturbations. Becuse of non-linearities, the effects of the background to the propagations of the perturbations (top-down) and the contributions of the propagations to the background itself (bottom-up) are non trivial problems.}
\label{fig:1}       % Give a unique label
\end{figure}

On the other hand, the top-down problem is concerned with whether the global state of the Universe can have effects on local systems, changing their dynamics. As we have seen, the mere identification of a local gravitational system is problematic in an expanding Universe. Spacetime subsystems are well defined, in a four-dimensional sense, when we can distinguish, globally, two consistent sets of properties within it. This is the case of a black hole,  defined as the non-causally connected part of the entire spacetime (impossibility of escaping to future null infinity). When spacetime is expanding, global notions are hard to establish, so we need quasi-local descriptions of a system (Ashtekar and Krishnan 2004).  For instance, in an expanding universe, it is hard to find black hole solutions. Finally, in the Newtonian approximation, the expanding Universe implies a change in the local inertia of the system that accelerates or deaccelerates bodies (Carrera and Giulini 2010). Structures of parsec scale such as galaxy clusters might be affected by this effect and have an observable change in their size.

All this interconnectedness shows something fascinating: understanding our surrounding might give us information about the entire Universe, and the other way around. A typical issue of emergence arises at this point. If the Universe is unique, we cannot distinguish boundary conditions from physical laws locally (Ellis, 2002). If the ``global physical laws'' are different, as the Universe is changing, we might observe changes in our ``local physical laws'', for instance, by variations of the fundamental constants of nature.

\section{Conclusions}

Spacetime is material. In a sense that we discussed above, it can change and interact. But it is a different kind of matter. In this chapter, we discussed some of the issues regarding the materiality of classical spacetime. The radical novelty introduced with General Relativity, i.e. background independence, has profound consequences on the ontological status of what we understand as time, space, and matter. This makes the characterization of some of its material properties conceptually difficult to establish. Following a clear materialistic ontology, we discussed the philosophical issues of defining self-interaction, spacetime parts, spacetime energy, and the concept of scale within spacetime. An understanding these issues is necessary to get a clear picture that can lead the way to a quantum theory of gravity, if such a theory could be formulated.

\begin{acknowledgement}
I would like to dedicate this contribution to the memory of Mario Bunge. I'm grateful to G.E. Romero for his intelectual guidance, and to Federico Lopez Armengol for many discussions on the nature of spacetime. This work was partially funded by a CONICET fellowship.
\end{acknowledgement}

\section*{Appendix: dictionary of technical terms}
%\addcontentsline{toc}{section}{Appendix}
In this section we give some useful definitions regarding spacetime and spacetime symmetries to complement the article.

\begin{description}
	\item[\textbf{Spacetime model}] A spacetime model $M= (\mathcal{M},\mathbf{g})$ is a real, four-dimensional connected $C^{\infty}$ Hausdorff manifold with no boundaries, with a globally defined $C^{\infty}$ tensor field $\mathbf{g}$ of type (0,2), non-degenerate, and Lorentzian (Wald 1984).
\end{description}

\begin{description}
	\item[\textbf{Affine connection}] A covariant derivative in the direction of a vector $\mathbf{W}$, $\nabla_{\mathbf{W}}$, is a derivative operator that transforms tensors into tensors. The operator is completly determined through its aplication to a basis of spacetime $\mathbf{e}_a$, in the direction of the same basis 
	\begin{equation}
	\nabla_a \mathbf{e}_b = \omega^{c}_{ab} \mathbf{e}_c.
	\end{equation}

The quantity $ \omega^{c}_{ab}$ is called the affine connection of $\nabla$.
\end{description}

\begin{description}
	\item[\textbf{Equation of motion}] Following Giulini, 2007, we represent a general equation of motion (EOM) of a given physical theory by:
\begin{equation}
	E[g,\gamma, \Phi; \Sigma] = 0,
\end{equation}

where $E$ is some differential operator, $g$ represents the spacetime metric, $\gamma$ represents particle and $\Phi$ represents physical fields, and $\Sigma$ represents some geometrical and non-dynamical structures that must be given. The latter are usually referred as the \textit{absolute structures} of the theory and can be referential or non-referential. There is no general consensus about a formal definition of the absolute structures of a given theory. The dynamical fields $\left(g,\gamma, \Phi\right)$ are unknown and meant to be solved, given $\Sigma$.
\end{description}

\begin{description}
	\item[\textbf{Covariance}] An EOM is \textbf{covariant} under the subgroup $G \subseteq \mathrm{Diff}(M)$ iff for all $f\in G$:
    \begin{equation}
    	E[g,\gamma, \Phi; \Sigma] = 0 \Longleftrightarrow E[f \cdot g,f \cdot \gamma, f \cdot \Phi; f \cdot \Sigma] = 0.
    \end{equation}
\end{description}
\textit{Remark}: As stated by Giulini, covariance requires the equation to `live on the manifold'. In other words, to refer to well-defined geometrical objects with given transformation laws under $f \in G$. The equation remains valid after the action of $f \in G$, but it is different from the original, since the components of $\Sigma$ are different.

\begin{description}
    \item[\textbf{Invariance}] An EOM is \textbf{invariant} under the subgroup $G \in \mathrm{Diff}(M)$ iff for all $f\in G$:
    \begin{equation}
    	E[g,\gamma, \Phi; \Sigma] = 0 \Longleftrightarrow E[f \cdot g,f \cdot \gamma, f \cdot \Phi; \Sigma] = 0.
    \end{equation}
\end{description}
\textit{Remark}: An invariant diffeomorphism $f \in G$ keeps the equation identical, since the components of $\Sigma$ are unchanged. In this way, invariance is much more restrictive than covariance. The invariant group of a given equation is inherited by the invariance group of its absolute structures $\Sigma$. Since the transformed equation is identical to the original, from a solution of the dynamical fields we can obtain a whole set of solutions related by invariant diffeomorphisms.

\begin{description}
    \item[\textbf{General Covariance}]  An EOM is \textbf{general covariant} if it is covariant under the group $G = \mathrm{Diff}(M)$.
\end{description}

\textit{Remark}: The equations of any theory can be written, in principle, in a general covariant way (Rovelli 2004). This process is usually carried out to make the equations of a given theory compatible with GR, but this is not necessary the case. In fact, a given covariant theory under an specific group of transformation can be made general covariant in many different ways. The key assumption in general relativistic covariantization is that the metric is the only external geometrical object, and thus, the connection that it is introduced to define a meaningful notion of derivation is the Levi-Civita connections. Adding, for instance, additional tetrad fields or other geometrical objects to construct a more general connection would render a Lorentz covariant equation general covariant but with the wrong limit in GR. Related to this is the notion of

\begin{description}
    \item[\textbf{General Invariance}]  An EOM is \textbf{general invariant} if it is invariant under the group $G = \mathrm{Diff}(M)$.
\end{description}

\textit{Remark}: As we pointed out earlier, a precise definition of background independence is very subtle and is not equal to general invariance. We shall present the classical definition by Anderson, 1976 that is not, however, free of problems (see also Friedman, 1973):

\begin{description}
	\item[\textbf{Background Independence} (Anderson)] A theory is background independent if its dynamical EOM are free of absolute structures.
\end{description}

\textit{Remark}: From the previous definitions, it follows that the equations of motion of any background independent theory, written in a general covariant way, are general invariant as well. That is the case of GR. However, let us note that there are various aspects in GR that can be consider absolute structures, e.g. dimensions and signature, and others known geometrical objects, as volume elements and vector fields.

%\bibliography{references}

\end{document}